\documentclass{easychair}

\usepackage{amsmath}
\usepackage{amssymb}
\usepackage{booktabs}
\usepackage{graphicx}
\usepackage{xcolor}

\usepackage{url}
\newcommand{\lean}{\begingroup
  \urlstyle{tt}%
  \def\UrlBreaks{\do\.\do\_\do\-}%
  \def\UrlBigBreakPenalty{0}\def\UrlBreakPenalty{0}%
  \Url}

\title{ProofJudge: Tool-Grounded LLM Evaluation of Formal Proof Quality in Mathlib}

\titlerunning{ProofJudge: Tool-Grounded LLM Evaluation of Formal Proof Quality in Mathlib}
\authorrunning{}

\author{
  Shane Caldwell
}

\institute{
  Dreadnode\\
  \email{shane@dreadnode.io}
}

\begin{document}

\maketitle

\begin{abstract}
Formal proofs in Lean~4 that pass the kernel's type checker can nonetheless vary widely in quality.
We introduce ProofJudge, an agentic LLM-as-judge system that scores formal proof quality along five dimensions beyond correctness: library leverage, automation fit, structural clarity, statement quality, and Mathlib conventions.
We evaluate ProofJudge on a novel dataset of 218 declarations drawn from distinct Mathlib PRs.
The judge agent is grounded by tool access to the commit the PR is applied to, enabling it to query the library state when scoring.
A judge is considered aligned with human preferences when it rates the version of the PR Mathlib accepted above the initial version that was sent back for revision. 
All six judge models evaluated recover the reviewers' preference well above chance, from 80.8\% to 63.5\%, and two open-weight judges reach roughly 70\% at a tenth of the best judge's cost.
We release the judge harness,\footnote{\url{https://github.com/SJCaldwell/ProofJudge}} evaluation dataset,\footnote{\url{https://huggingface.co/datasets/SJCaldwell/proofjudge}} and evaluation traces\footnote{\url{https://huggingface.co/datasets/SJCaldwell/proofjudge-eval-traces}} as open-source artifacts to support further research.
\end{abstract}

\section{Motivation and Background}

As reinforcement learning from verifiable reward has become popular for eliciting skills in language models\cite{lambert2025tulu3pushingfrontiers}, Lean~4's ability to act as a verifier for mathematical proofs has emerged as a natural way to provide reward, leading to increased mathematical proof writing capabilities, both with human collaboration\cite{song2025leancopilotlargelanguage} and autonomously\cite{ren2025deepseekproverv2advancingformalmathematical}. 
While the proofs that pass the type checker are provably correct, the question remains: what is the value of these generated proofs for human mathematicians\cite{avigad2026mathematiciansageai}? Critics argue that the value of formalization is not in proving the correctness of a statement, but rather better understanding the results and ``building libraries and infrastructure to support future work''. 
Mathlib, the largest Lean~4 mathematical library, enforces quality standards through human code review: reviewers evaluate tactic hygiene, generality of lemma statements, proof structure, naming conventions, and---most critically---whether a proof decomposes into independently reusable components that extend the library's API surface.
These deeper structural and stylistic properties~\cite{van_Doorn_2020} are often violated in current LLM-generated proofs, creating a heavy burden on skilled Mathlib reviewers as the cost of generating valid proofs is reduced year over year.
It remains to be seen whether LLMs could perform this review role necessary to keep library quality high as the scale of automated proof writing increases in the future.
\section{Method}

We introduce \textbf{ProofJudge}, an Agentic Judge~\cite{you2026agentasajudge} system that evaluates formal proof quality beyond compilation. 
The Lean kernel provides an objective correctness anchor, while the judge evaluates softer quality dimensions---for example, statement quality---via a rubric based on Mathlib's review norms. 
ProofJudge's tool access (querying Mathlib via bash) enables the judge to ground its assessments in the actual state of the library as a human reviewer does.
\subsection{Rubric Design}

Recent work applies LLM judges to the semantic correctness of PRs\cite{xin2026apebenchevaluatingautomatedproof}; we instead focus on whether a proof meets the qualitative standards for library inclusion.

The rubric asks the judge agent to decompose the verdict: the model scores five dimensions independently on a 1--10 scale.
Those are: library leverage, automation fit, structural clarity, statement quality and Mathlib conventions.
The harness takes those numbers and weights them towards a final score.

\subsection{Dataset Construction}

We evaluate ProofJudge on an initial dataset of 218 declaration pairs drawn from Mathlib pull requests. 
To develop the dataset, we used \texttt{claude-sonnet-5} to review past PRs and determine which had significant differences between the earliest and final revision that did not come down to linting or involve a new declaration. 
The dataset, which we have open-sourced, includes a dev split of 123 declaration pairs that was used to tune the rubric, along with the 218-pair test split for the eval itself.
\section{Results}

For each pair, the judge scores both the 
pre-revision and post-revision proof independently, and we measure whether the judge's preference aligns with the reviewer's. 
By alignment, we refer to the post-revision proof (that was accepted into the library) receiving a higher score than the pre-revision proof (that was rejected). 
The judge considers each PR independently, and is unaware of the score it provided the other revision or that any other revision is being scored at all. 
The judge agent is allowed to use twenty tool calls before it is forced to make a determination to limit inference costs.

We evaluate six judges---three open-weight, three closed---on the 218-pair test split with three replicates each. 
We compare against a 50\% random-chance baseline. 

\begin{table}[h!]
\centering
\small
\setlength{\tabcolsep}{5.5pt}
\begin{tabular}{llcr}
\toprule
Judge & Weights & Alignment \% (95\% CI) & USD/pair \\
\midrule
\texttt{claude-sonnet-5}   & closed & 80.8 \ (73.6--88.2) & 1.392 \\
\texttt{gemini-3.7-flash}  & closed & 75.2 \ (69.3--81.2) & 0.186 \\
\texttt{muse-glimmer-30b}  & open   & 70.2 \ (63.4--77.3) & 0.140 \\
\texttt{inkling-small}     & open   & 69.3 \ (60.2--77.8) & 0.126 \\
\texttt{gpt-5.4-mini}      & closed & 68.7 \ (60.6--77.1) & 0.136 \\
\texttt{deepseek-v4-flash} & open   & 63.5 \ (56.0--71.6) & 0.029 \\
\bottomrule
\end{tabular}
\caption{Six judges on the 218-pair test split, three replicates each. Intervals are a clustered bootstrap over declarations and replicates. Cost is published list rates times token counts.}
\label{tab:judges}
\end{table}

\begin{figure}[h!]
\centering
\includegraphics[width=\textwidth]{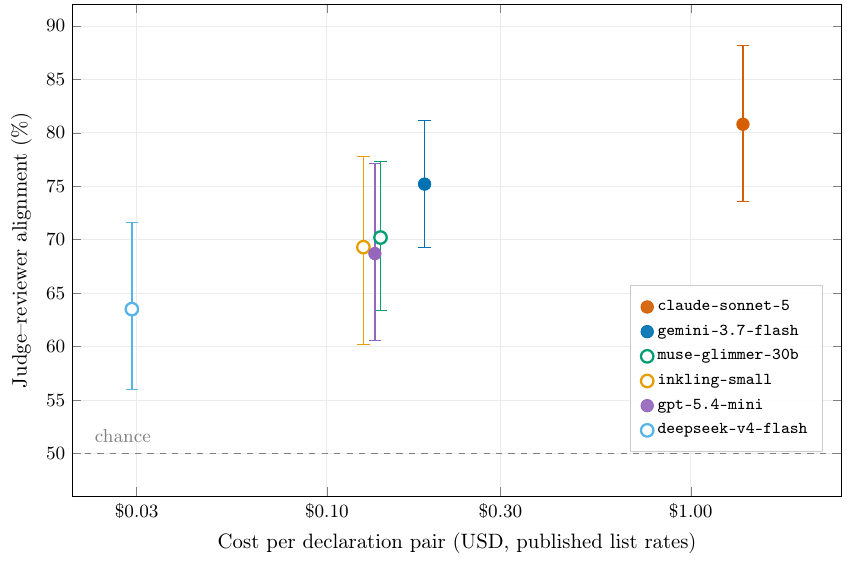}
\caption{Alignment against cost per declaration pair, with the intervals of Table~\ref{tab:judges}.
Open-weight judges are drawn hollow.}
\label{fig:pareto}
\end{figure}

Every judge recovers the reviewers' preference far above chance, from \texttt{claude-sonnet-5} at 80.8\% of declarations down to \texttt{deepseek-v4-flash} at 63.5\%, each at $p<10^{-5}$ by sign test. Whatever reviewers apply when they send a proof back is legible to a model reading the two revisions.

Reviewing Mathlib PRs this way is not limited to frontier models.
Two open-weight judges, \texttt{muse-glimmer-30b} and \texttt{inkling-small}, sit at 70.2\% and 69.3\%.

However, the judges are noisy.
Re-running one over the same declarations flips between a fifth and nearly half of its verdicts, and a single run's interval hides that badly: the same ablation returns $p=0.34$ on one replicate and $p=0.006$ on another.
A single-run version of this judge benchmark will report differences that replication does not support.

\subsection{An Example Grading}
To help illustrate the eval, we provide a representative example of a PR that was evaluated correctly during the ProofJudge evaluation.

In PR~\href{https://github.com/leanprover-community/mathlib4/pull/11640}{11640}, the initial proof of \lean{Set.restrictPreimage_isClosedMap} reconstructed a result that Mathlib already provided. 
The merged revision replaces it in one line, with \lean{H.restrictPreimage} \texttt{s}.
Scoring the initial variant, the judge searched the repository for the existing declaration, quoted the signature it found, and marked the initial PR down on the dimension of library leverage.
Every judge reduced the score of the initial PR in every rollout.
Note that while this would not be possible with regular expressions or linting, an agent with tool access can search the library, understand the semantic meaning of what it finds, and act as a reviewer would.

\section{Conclusion}

Correctness is not the only property of formal proof that matters.
It is important to mathematicians that proofs be understandable, and that contributions to mathematical libraries enable future work and remain accessible to readers.
While there is work to be done, the alignment scores from modern models evaluated with ProofJudge suggest that these qualities are not illegible to language model agents. 
With that established, this signal should be used to relieve maintainer burden where possible, and create agents capable of writing Lean 4 PRs that maintainers would welcome.

\subsection{Future Work}

Future work should focus on making judges that perform at the same level as \texttt{claude-sonnet-5} but cheaper and open-source to help reduce undue burden on Mathlib maintainers.
In addition to a cost focus, a less noisy judge would give a model a signal to iterate a PR against, improving quality before a human reviews it.
Reinforcement learning using the ProofJudge results as a reward may create models that write better proofs before review.

Two directions for reducing judge noise are left to future work: splitting the rubric into five single-dimension judges so each makes more focused use of tools, and treating the tool-call budget itself as a tunable parameter.


\bibliographystyle{plain}
\bibliography{proofjudge}

\end{document}